# An open dataset of article processing charges from six large scholarly publishers (2019-2023)

Leigh-Ann Butler[1,2], Madelaine Hare[2,3], Nina Schönfelder[4], Eric Schares[2,5], Juan Pablo Alperin[2,6] & Stefanie Haustein[2,3,7*]

[1] University of Ottawa Library, Ottawa (Canada)
[2] Scholarly Communications Lab, Ottawa/Vancouver (Canada)
[3] School of Information Studies, University of Ottawa, Ottawa (Canada)
[4] Bielefeld University Library, Bielefeld (Germany)
[5] University Library, Iowa State University, Ames (USA)
[6] Simon Fraser University, Vancouver (Canada)
[7] Centre interuniversitaire de recherche sur la science et la technologie (CIRST), Université du Québec à Montréal, Montréal (Canada)
[*] *Corresponding author:* stefanie.haustein@uottawa.ca

## Abstract
This paper introduces a dataset of APCs produced from the price lists of six large scholarly publishers – Elsevier, Frontiers, PLOS, MDPI, Springer Nature and Wiley – between 2019 and 2023. APC price lists were downloaded from publisher websites each year as well as via Wayback Machine snapshots to retrieve fees per journal per year. The dataset includes journal metadata, APC collection method, and annual APC price list information in several currencies (USD, EUR, GBP, CHF, JPY, CAD) for 8,712 unique journals and 36,618 journal-year combinations. The dataset was generated to allow for more precise analysis of APCs and can support library collection development and scientometric analysis estimating APCs paid in gold and hybrid OA journals.

## 1. Introduction
This paper introduces an open dataset (Butler et al., 2024) of article processing charge (APC) prices collated from publisher price lists which can be used for scientometric study, analyses of the scholarly publishing market, or for library collections management. This dataset complements the Butler et al.'s (2022) dataset, first used and described in Butler et al.'s (2023) study, by capturing annual APC prices beyond 2018 and including two large multidisciplinary exclusive OA publishers (Frontiers and MDPI).

In the following sections, we describe the methods used to collect, process, and clean the APC list prices to generate a coherent and clean dataset that comprises annual OA fees for gold and hybrid journals published by Elsevier, Frontiers, MDPI, PLOS, Springer Nature and Wiley from 2019 to 2023. We then summarize and discuss the dataset at a high level, by presenting preliminary analysis and observations. We conclude by addressing potential uses of this open dataset.

## 2. Data and methods

### 2.1 Data Sources

The dataset combines and standardizes data from the APC price lists of six large publishers (i.e., Elsevier, Frontiers, MDPI, PLOS, Springer Nature, and Wiley). The dataset includes APC prices for 8,712 unique journals and 36,618 data points, so-called journal-year combinations spanning five years (2019-2023). The APC data was imported from a total of 37 publisher price lists and cleaned to produce one coherent and reusable dataset.

APC prices were captured from several sources and in various formats. Annual price lists (2019-2023) were regularly downloaded from publisher websites by one of the authors (NS). These price lists were typically provided in downloadable PDFs, structured XLSX files, or displayed as HTML on publisher websites and contained information such as ISSNs, journal title, OA status, APC list price, and currency. When price lists were not made available on publisher websites in a downloadable format (such as with Frontiers), APCs were scraped from individual journal web pages and collated into one XLSX file for import.

We selected the price list that was published or collected in or around June of each year—the time of year for which we were able to most consistently identify a price list from the collection retrieved by NS. We argue that an APC list price collected mid-year would be most representative of the actual fee paid in the publication year by equally over- or underestimating fees paid in the first vs second half of each year. However, using a fee retrieved mid-year is different from Delta Think's approach, which is a commercial service that collects price lists from publishers each January to distribute aggregated APC data to paying subscribers (Pollock & Staines, 2024). Compared to Delta Think's approach, our dataset is available openly.

### 2.2 Selection of files, data cleaning, and metadata enrichment

We used a mix of manual and automatic methods for combining the downloaded files into the main dataset. PDF files from Elsevier were in tabular form and were copy-pasted into an XLSX file, concatenated together by page. We automated the process for Wiley and Springer Nature by using a Python script to reorganize the data from the XLSX files. Columns were then mapped to the headings found in the final dataset. The same procedure of column mapping was done with files from PLOS and MDPI. The year 2023 for Frontiers was scraped from each individual journal webpage as no single downloadable file was available. Frontiers provided a list of journal titles and APCs on their website for 2019-2022, which were captured as images, and were therefore not in machine-readable format. These were typed manually into an XLSX file for each year, and then imported into the APC dataset. The data source and file formats for price lists per publisher as well as the initial of the co-authors who collected the data are presented in Table 1.

Table 1. Overview of price lists, data sources, formats, and collectors

| Publisher | Data source(s) | Format(s) collected | Collector(s) |
|---:|---|---|---|
| Elsevier | Publisher website | PDF, xlsx | NS |
| Frontiers | Publisher website; Journal web pages; Wayback machine | HTML | NS; ES |
| MDPI | Wayback machine | HTML | ES |
| PLOS | Publisher website | HTML | NS |
| Springer Nature | Publisher website | xlsx | NS |
| Wiley | Publisher website | xlsx | NS |

We then cleaned the data, such as merging variant spellings of journal titles, and added key missing information, such as ISSNs, to produce v1 of this dataset which is available openly for reuse as Butler et al. (2024) on the Harvard Dataverse.

### 2.2.1 Unique identifiers

As the ISSN is not a unique identifier for a journal in the sense that a journal has multiple ISSNs (see 2.2.2), we assigned an internal unique identifier (ID) to each journal title that ranges from 1 to 9010 and merged spelling variations of the same journal, if identified. In some cases, we identified and cleaned discrepancies, such as two unique IDs being assigned to variant spellings of journal titles. The numbers are for dataset internal identification only and do not correspond to external identifiers.

### 2.2.2 ISSNs

Most publisher price lists include at least one ISSN, but do not always specify whether the identifier provided corresponds to the print (pISSN) or electronic (eISSN) serial number. Therefore, the dataset does not make such a distinction and instead has two ISSN columns (ISSN_1 and ISSN_2). In cases where publishers provide a single ISSN, we assign it to ISSN_1; in cases where publishers distinguish between eISSN and a pISSN, we assign the eISSN to ISSN_1 and the pISSN to ISSN_2. All ISSNs were standardized to include a hyphen after the first 4 digits. Through a validation process, we identified 117 cases where ISSNs were incorrectly attributed to journals for Wiley (n=76), Springer Nature (n=38), and Elsevier (n=3); these corrections were documented in the data cleaning notes file that accompanies the dataset. We confirmed the misattribution of ISSNs to publisher error in their price lists. Correct ISSNs (when found) were located from journal websites and the ISSN portal and updated for these particular entries in the dataset. We manually checked and corrected all issues so that an ISSN was only assigned to one unique journal ID.

### 2.2.3 Changes in publisher and OA status

In some cases, more than one publisher was associated with the same journal. This was usually due to journals transferring between publishers. A transfer was usually communicated via an announcement on a journal, publisher, or society website, so we made note of this in the comment column. A journal that transferred from one publisher to another will still have the same unique ID. In rare cases, the transferred journal was listed in the annual price lists of both publishers, which resulted in our dataset containing two APCs for the same year, one for each publisher. For example, our dataset contains two fees for *Acta Mathematica Scientia* (unique_id=15) in 2020 because the journal transferred from Elsevier to Springer. Similarly, if a journal flipped OA status (e.g., from hybrid to gold or vice versa), we have two APCs per unique ID per year, one for hybrid and one for gold. For example, Springer Nature's *Journal of Materials Science: Materials in Medicine* (unique_id=5723) and Wiley's *Ecography* (unique_id=7465) both flipped from hybrid to gold in 2021. Changes in publisher (n=11 journals) or OA status (n=2 journals) are the only situations in which we have two APCs per journal per year. In all other cases, each journal just has one APC per year.

### 2.2.4 Journal title variations

In some cases, multiple entries appeared in a year for the same journal based on name variations. In these cases, duplicate entries were removed, with the entry using the journal's title proper (as noted on publisher websites) kept in the dataset. We cleaned the dataset for spelling variations in journal titles (e.g. titles that were missing apostrophes or "the" preceding their title), matching titles by ISSN across all years. We noted spelling variations occurring in publisher's price lists from one year to another with inconsistencies in particular for the use of diacritics (e.g., Spanish, French, German). A record of these changes can be found in the data cleaning notes file of the published dataset.

### 2.2.5 Currency conversion

Publishers provided APCs in different currencies, with many publishers providing fees in multiple currencies including USD, EUR, GBP, JPY and CHF. We therefore decided to keep APCs in all currencies provided. To allow analysis across journals (and in different currencies), we filled gaps by converting between currencies using respective annual average conversion rates retrieved from ofx.com. These annual conversion rates are provided in the dataset (Butler et al., 2024). USD was the most frequently provided currency at 92% of the 36,618 journal-year combinations. The remaining 8% without USD original fees were converted from APCs provided in CHF, EUR or GBP. For each currency we specify in the respective column (e.g., APC_USD-originalORconverted) whether the provided APC was retrieved originally from the publisher (i.e., "original) or converted from a different currency (e.g., "converted from CHF"). Although no publisher provided APCs in CAD, we also provided converted fees in CAD to allow for immediate analysis of APC spending in the Canadian context.

### 2.2.5 Machine readability

To ensure the dataset was machine-readable, we separated variables with mixed data types into two variables. For example, some publishers included comments such as "see website" in the APC column. To ensure easy processing and machine readability, we converted such mixed data types into two columns. For example, we added the variable "APC_provided" to indicate "no" when publishers did not list a fee. For all rows with an APC provided, the APC columns for various currencies contain a value. In all other cases, the APC columns are Null. In addition, we provided the variable "APC_order", which indicates the order of APCs per journal per year. In case the publisher or OA status changed and there were two APCs per year, a value of 2 indicates that this is the APC associated with the publisher or OA status that the journal changed to. This is the case in 13 out of the 36,618 journal-year combinations. In these 13 cases for which the publisher or OA status changed during a year, the previous APC value is indicated with a value of 1. If we did not have more than one APC per journal per year because the journal did not change publisher or OA status during the year, the APC order has a value of 1. Programmatically, if one wishes to only keep one APC per journal per year, we recommend using the maximum value for the "APC_order", that is a value of 2 in case of a change and 1 if there was no change. A list of variables in the dataset with a brief description and example value for each is shown in Table 2.

**Table 2.** Variables in the APC dataset.

| Variable name | Description | Example value |
| --- | --- | --- |
| unique_id | Unique ID per journal. | 15 |
| Publisher | The publisher of the journal in the particular year. | Springer Nature |
| ISSN_1 | The first ISSN of the journal, usually but not always the electronic ISSN. | 1572-9087 |
| ISSN_2 | The second ISSN of the journal; usually but not always the print ISSN, if provided. | 0252-9602 |
| Journal | The proper title of the journal. | Acta Mathematica Scientia |
| OA status | The open access status of the journal, indicated by the publisher or journal website or price list. | Hybrid |
| APC_provided | Information on whether an APC value was provided in the price list. | yes |
| APC_order | Order of APCs when there are two APC values per year in case of publisher or OA status change. | 2 |
| APC_currency* <br> * for each of following currencies: USD, EUR, GBP, JPY, CHF, CAD | The article processing charge value as provided by the publisher or converted into each respective currency. | 2750 |
| APC_currency_original ORconverted | This field indicates whether the respective currency originally derived from publisher price lists or converted from another currency using an annual conversion rate. | original |
| APC_date | The date of collection or Waybackmachine snapshot of the published pricelist from which the APC value was taken. | 2020-10-22 |
| APC_year | The year of collection or Waybackmachine snapshot of the published pricelist from which the APC value was taken. | 2020 |
| APC_source | Original source of price list. | Publisher website |
| Collector | Name of person who collected the price list. | N. Schoenfelder |
| Comment | Further contextual or significant details. | Journal transferred from Elsevier to Springer Nature in 2019 |

## 3. Descriptive statistics of the dataset

In this section we present a statistical summary of the APC dataset, including the unique number of journals per publisher per year (Table 3) as well as an analysis of the list prices, including distributions per publisher, year and OA status, as well as annual trends. The dataset contains a total of 8,712 unique journal IDs and demonstrates a steady growth of titles per year from 6,643 in 2019 to 7,985 unique titles in 2023. Note that the total number across all years exceeds the total in 2023. This is due to some titles no longer being included in price lists in later years. This could be due to several reasons, including the journal switching to a publisher that is not included in our dataset, a journal no longer offering OA publishing options, or a journal ceding publication.

Table 3 also demonstrates the size differences in portfolios between publishers. Elsevier (n=3,094), Springer Nature (n=2,952) and Wiley (n=2,018) clearly lead with the largest APC portfolios, which can be explained by their large number of paywalled journals that offer hybrid OA publishing options. Frontiers (n=221), MDPI (n=431) and PLOS (n=12) publish considerably fewer journals, but all of their titles are fully OA (i.e., gold).

**Table 3.** Number of unique journal IDs per publisher and per year

| Year | Elsevier | Frontiers | MDPI | PLOS | Springer Nature | Wiley | All publishers |
|---|---|---|---|---|---|---|---|
| 2019 | 2,260 | 61 | 205 | 7 | 2,568 | 1,542 | 6,643 |
| 2020 | 2,292 | 80 | 272 | 7 | 2,691 | 1,610 | 6,952 |
| 2021 | 2,672 | 108 | 367 | 12 | 2,715 | 1,659 | 7,533 |
| 2022 | 2,550 | 148 | 405 | 12 | 2,723 | 1,667 | 7,505 |
| 2023 | 2,704 | 221 | 427 | 12 | 2,725 | 1,896 | 7,985 |
| all years | 3,094 | 221 | 431 | 12 | 2,952 | 2,018 | 8,712 |

There is a total of 1,110 journal-year combinations with missing APC information, which are attributed to Springer Nature (n=712), Elsevier (n=365), MDPI (n=32) and Wiley (n=1). Through a manual check of the websites of a sample of these journals with missing values we noted a few different reasons for these gaps, which included temporary APC waivers or an academic society paying fees. For Elsevier we also found 204 journal-year combinations for which an OA status was not provided, 16 of which with an APC listed (Table 4). In a future update of the dataset, we plan to manually retrieve missing APCs and OA status information.

**Table 4.** Descriptive statistics of APCs per publisher and OA status (2019-2023)

| Publisher | OA status | Number of journal-year combinations | APC in USD | | |
|---|---|---:|---:|---:|---:|
| | | | Min | Max | Average |
| all publishers | | 36,618 | 0 | 11,690 | 2,859 |
| | Gold | 8,499 | 0 | 8,900 | 1,977 |
| | Hybrid | 26,993 | 0 | 11,690 | 3,137 |
| | No OA status provided | 16 | 1,250 | 3,400 | 2,288 |
| | No APC provided | 1,110 | | *n/a* | |
| Elsevier | | 12,478 | 150 | 10,100 | 2,736 |
| | Gold | 2,515 | 200 | 8,900 | 1,891 |
| | Hybrid | 9,582 | 150 | 10,100 | 2,959 |
| | No OA status provided | 16 | 1,250 | 3,400 | 2,288 |
| | No APC provided | 365 | | *n/a* | |
| Frontiers | Gold | 618 | 0 | 3,295 | 2,093 |
| MDPI | | 1,676 | 0 | 2,895 | 1,383 |
| | Gold | 1,676 | 0 | 2,895 | 1,383 |
| | No APC provided | 32 | | *n/a* | |
| PLOS | Gold | 50 | 1,595 | 6,300 | 2,740 |
| Springer Nature | | 13,422 | 0 | 11,690 | 3,041 |
| | Gold | 2,312 | 0 | 6,850 | 2,348 |
| | Hybrid | 10,398 | 0 | 11,690 | 3,195 |
| | No APC provided | 712 | | *n/a* | |
| Wiley | | 8,374 | 0 | 6,100 | 3,106 |
| | Gold | 1,360 | 0 | 5,740 | 2,139 |
| | Hybrid | 7,013 | 950 | 6,100 | 3,294 |
| | No APC provided | 1 | | *n/a* | |

Of the 36,618 journal-year combinations with an APC value provided, 123 listed an APC of $0, which indicates that at the time of data collection authors did not have to pay a fee to publish OA in the respective journal. This could be due to temporary APC waivers for marketing purposes or permanent waivers if the publication costs are covered by a third party (i.e., diamond OA journals). While 122 of these journals were gold OA journals, we also noted that the *International Journal of Obesity Supplements* published by Springer Nature was listed as a hybrid journal with an APC of $0.

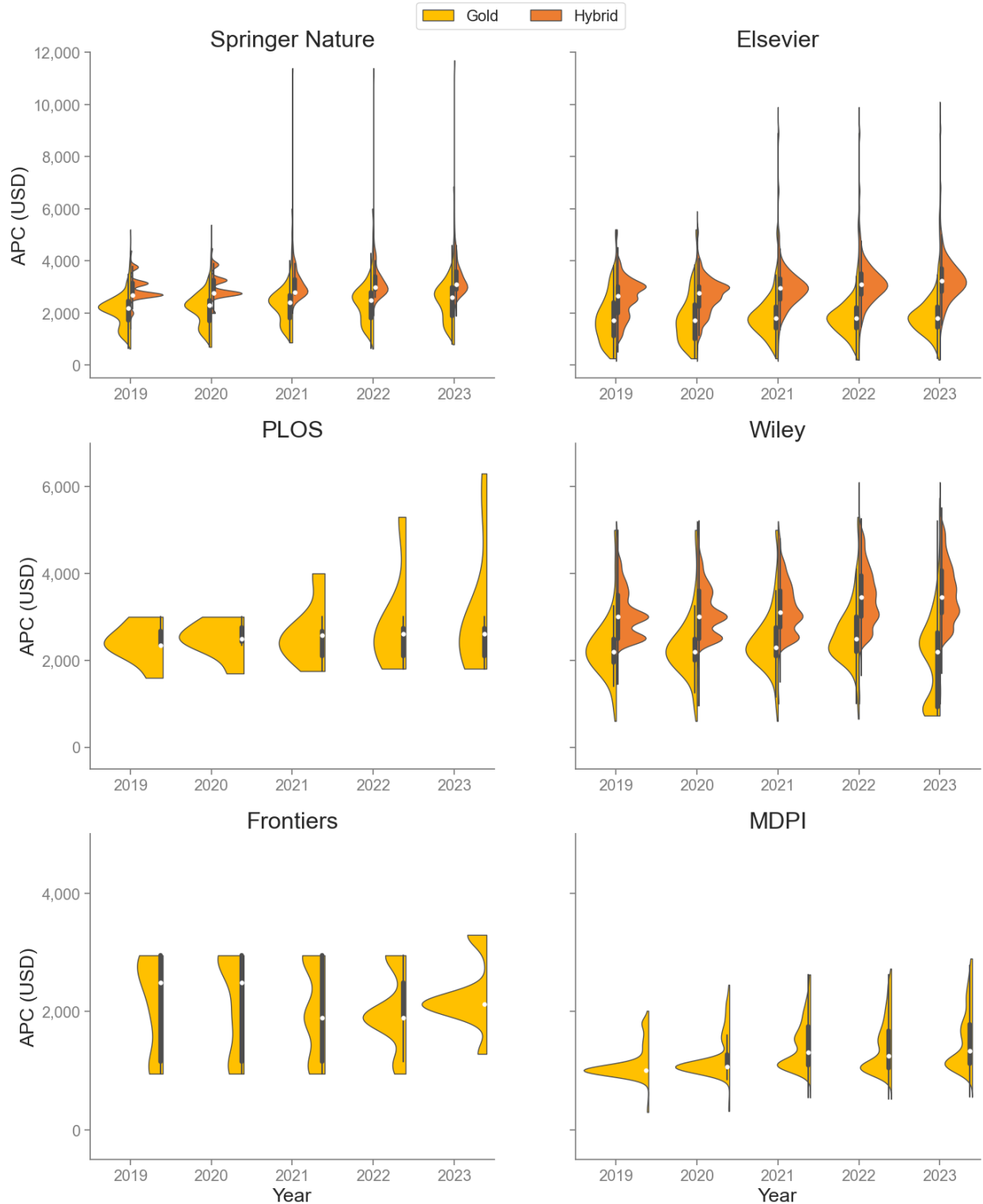

**Figure 1.** Violin plot showing the number of annual APCs (USD) per journal per publisher per OA status, 2019-2023. 123 journal-year combinations with an APC of $0 were excluded.

For the remaining journal-year combinations with an APC above $0, fees ranged from $150 for Elsevier's *Materials Today: Proceedings* between 2019 and 2021 to $11,690 for 37 *Nature* journals in 2023. The average APC per paper for all publishers and years was $1,977 for gold OA and $3,137 for hybrid OA (Table 4). This corroborates extensive evidence from the literature that average hybrid fees significantly exceed gold fees, even if journals with hybrid OA options are already financed through annual subscription fees (Bakker et al., 2017; Björk & Solomon, 2014; Butler et al., 2023; Pinfield et al., 2016; Solomon & Björk, 2012). The distribution of price points comparing gold and hybrid per publisher can also be seen in the violin plot in Figure 1. Comparing price points by publisher, average gold APCs were lowest for MDPI ($1,383) and highest for PLOS ($2,740). Average hybrid fees were very similar across the three publishers that offered them, ranging from $2,959 for Elsevier and $3,195 for Springer Nature to 3,294 for Wiley (Table 4, Figure 1). This reflects the trend set by Springer in 2004, when they introduced hybrid APCs at $3,000 through their OpenChoice program (Björk & Solomon, 2015; Copiello, 2020).

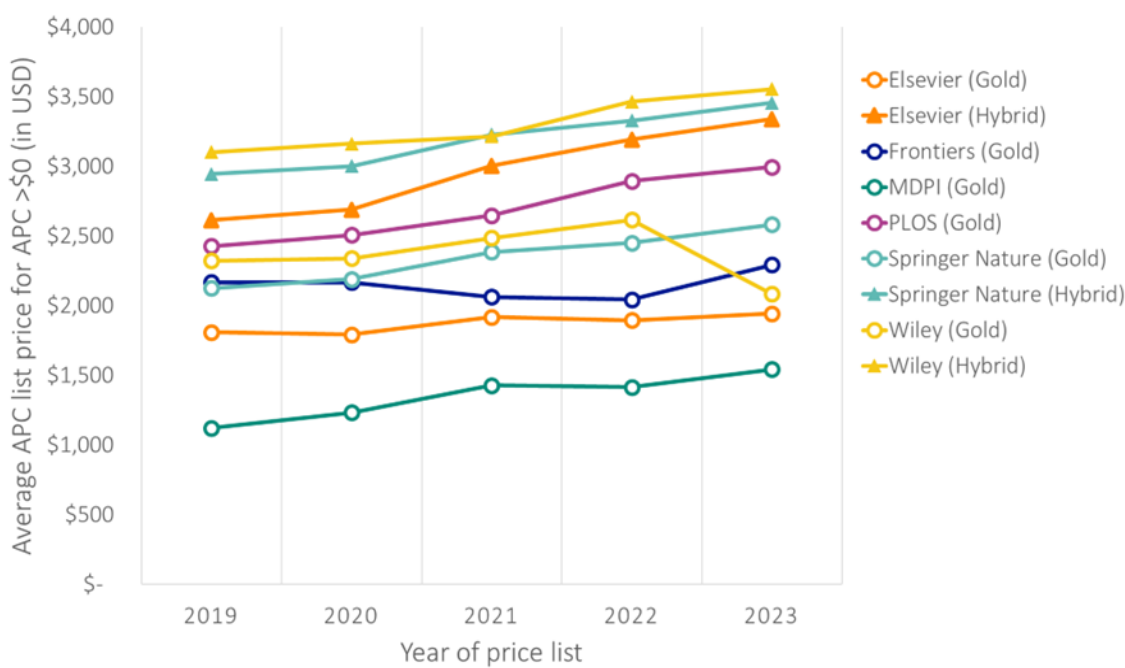

**Figure 2.** Average gold and hybrid APC list prices per publisher. 123 journal-year combinations with an APC of $0 were excluded.

The violin plots (Figure 1) as well as the average APCs in Figure 2, also demonstrate that over the period 2019-2023, most publishers increased APCs from year to year, with the exception of Wiley gold OA fees, which drop from an average of $2,615 in 2022 to $2,083 in 2023. This decline might be the result of Wiley's acquisition of lower-priced Hindawi OA journals (Wiley, 2021). Comparing OA only publishers Frontiers, MPDI and PLOS, the latter shows significantly higher average APCs. However, with 12 journals in 2023, PLOS publishes a significantly lower number of titles than Frontiers and MDPI, who had 427 and 221 gold OA journals in their portfolio in 2023, respectively. MDPI exhibits the lowest APCs among all publishers analyzed. For example, in 2019, 74% of MDPI journals charged a fee of $1,007 (=1,000 CHF). In 2021, MDPI began to increase fees on average to provide a larger range of price points. Their share of journals charging 1,000 CHF ($1,094 in 2021) decreased to 45%, and 19% charged an APC of 1,400 CHF (=$1,532). As shown in the violin plots in Figure 1, Frontiers demonstrates an almost opposite trend moving from fairly evenly distributed price points from 2019 to 2021 to a focus on $2,125 (63%) and some higher priced titles at $3,295 (19%) in 2023.

The Sankey diagram in Figure 3 analyzes APCs by comparing list prices in 2023 to those charged in 2019 on the journal level. For each journal for which an APC was provided in both years (n=5,842), we calculated the percent increase of the fee from the first to the last year in our dataset and categorized journals according to their 2019 to 2023 APC change as journals that decreased their APC (n=462), those that had the same APC in both years ("no change", n=180) and those that increased their APC, either within (n=2,877) or above (n=2,323) the 19% currency inflation from 2019 to 2023. Note that in the Sankey diagram we excluded the 15 journals that changed publisher. Overall, an overwhelming majority (89%, n=5,192) of journals increased their OA fees, including 40% which increased above the 19% inflation. 3% of journals did not change their APCs, and 8% lowered their fees from 2019 to 2023. Comparing publishers, MDPI increased all but one (100%) of their journals' OA fees above inflation, followed by Elsevier (57%) and Frontiers (52%). At 71%, Springer Nature and PLOS were the two publishers with the largest percentage of journals increasing APCs within inflation, followed by Wiley (58%) and Frontiers (48%). Among the six publishers analyzed, Elsevier was the one that most frequently lowered fees (14%) or kept them constant (7%).

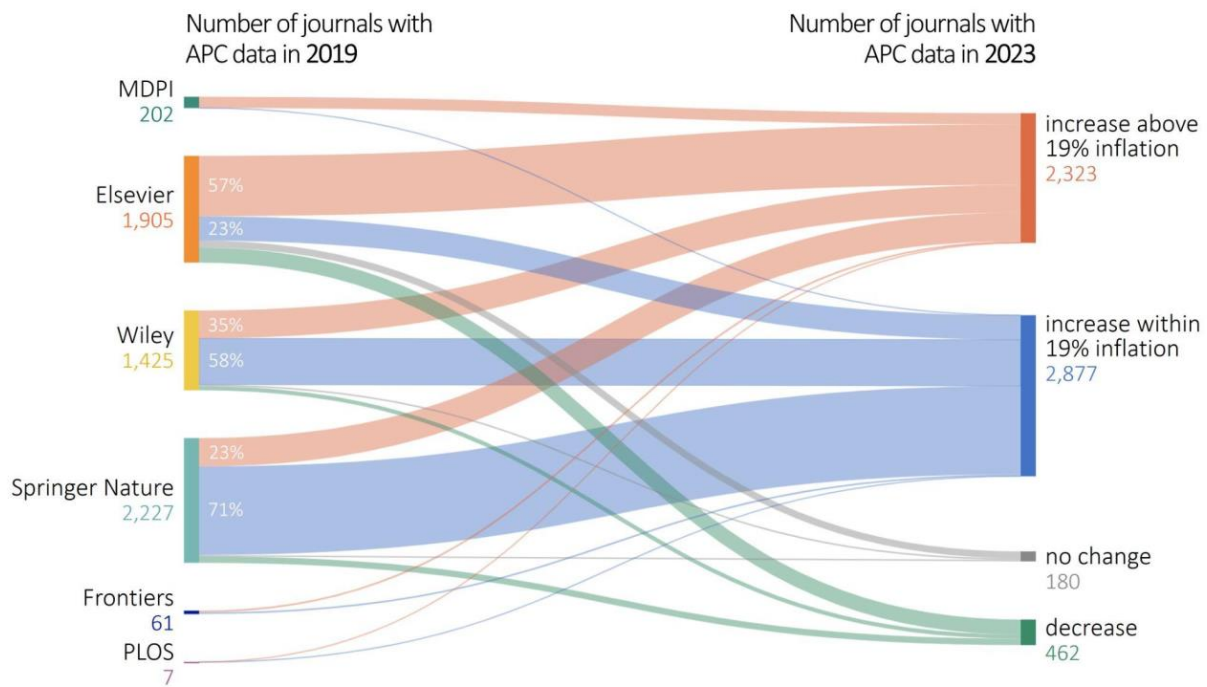

**Figure 3.** Comparison of APCs in 2023 vs 2019 per publisher for journals with APC data in both years (n=5,827). 15 journals which changed publishers were excluded from the visualization.

## 5. Limitations

This dataset possesses limitations relating to its methodological approach and the accuracy of its data. We relied on information provided by publishers in their price lists, potentially reproducing any inaccuracies. However, we considerably enhanced the provided data by generating a coherent and consistent dataset in machine-readable format. We focused on processing the data into a structured, reusable format with enriched and consistent metadata. This initial version of the dataset aims to reflect information provided by publishers with some enhancements of journal titles and ISSNs. Future versions of the dataset will include filling missing APCs and OA status from journal websites through a manual verification process. We also plan to identify diamond and delayed OA journals. Moreover, the dataset does not indicate waivers or discounts granted to individual authors since such information would only be available in the invoices issued by publishers or financial records of universities or funders who covered their fee. Despite these limitations, this dataset can be a valuable tool in estimating APC spend.

## 6. Conclusion and outlook

This dataset offers a range of potential applications for analyses, including, but not limited to, estimating APC spending, supporting collections development, or understanding the costs and value of read-and-publish agreements. While it can support professional practice in institutional libraries in such ways, from a research perspective this dataset can also aid in framing further potential areas of inquiry: it can provide an evidence base for all stakeholders engaged in scholarly publishing and serve as a foundation for quantitative science studies that investigate trends in the OA landscape and academic publishing market.

Future versions of the dataset will aim to supplement missing APC data and verify journal OA status through manual curation and verification. We will also seek to provide APCs from additional publishers and larger timespans. We aim to identify the subset of gold OA journals that could be considered diamond OA (Simard et al., 2024) and will flag journals that offer delayed OA. As we continue to build this dataset, careful consideration will be given to the evolving landscape, such as Wiley's decision to sunset and integrate the Hindawi brand into its OA portfolio (Wiley, 2023) and to close 19 journals due to questionable and fraudulent publication practices, including paper mills (Subbaraman, 2024).

Feedback on the dataset or data contributions can be sent by email to contact@scholcommlab.ca.

## Open science practices

The open dataset introduced and analyzed in this paper is available for download and reuse under a CC-0 license on the Harvard Dataverse at https://doi.org/10.7910/DVN/CR1MMV. A citation to the dataset is provided in the reference list as Butler et al. (2024).

## Acknowledgments

The authors would like to thank Dan Brockington for pointing us to price lists for MDPI and Chantal Ripp for support with research data management. We would also like to thank Poppy Riddle for creating an interactive website to explore the dataset which is available via scholcommlab.ca.

## Author contributions

## Competing interests

## Funding information

The authors did not receive funding for this project.